\documentclass[amssymb,aps]{revtex4b3}
\usepackage{graphicx}
\begin{document}
\draft
\title{Ferromagnetism in III-V and II-VI semiconductor structures}
\author{T. Dietl}
\email{dietl@ifpan.edu.pl}
\affiliation{Institute of Physics and
College of Science, Polish Academy of Sciences, \\ al.
Lotnik\'{o}w 32/46, PL-00-668 Warsaw, Poland and\\ Laboratory for
Electronic Intelligent Systems, Research Institute of Electrical
Communication,\\ Tohoku University, Sendai 980-8577, Japan}
\author{H. Ohno}
\email{ohno@riec.tohoku.ac.jp}
 \affiliation{Laboratory for
Electronic Intelligent Systems, Research Institute of   Electrical
Communication,\\ Tohoku University, Sendai 980-8577, Japan}

\begin{abstract}
The current status of research on the carrier-mediated ferromagnetism in tetrahedrally
coordinated semiconductors is briefly reviewed. The experimental results for III-V
semiconductors, where Mn atoms introduce both spins and holes, are compared to the
case of II-VI compounds, in which the ferromagnetism has been observed for the
modulation-doped p-type Cd$_{1-x}$Mn$_x$Te/Cd$_{1-y-z}$Mg$_y$Zn$_z$Te:N heterostructures,
and more
recently, in Zn$_{1-x}$Mn$_x$Te:N epilayers. On the theoretical side, a model is presented,
which takes into account: (i) strong spin-orbit and $kp$ couplings in the valence band;
(ii) the effect of confinement and strain upon the hole density-of-states and response
function, and (iii) the influence of disorder and carrier-carrier interactions, particularly
near the metal-to-insulator transition. A comparison between experimental and
theoretical results demonstrates that the model can describe the magnetic circular
dichroism, the values of $T_C$ observed in the studied systems as well as explain the
directions of the easy axis and the magnitudes of the corresponding anisotropy fields
as a function of confinement and biaxial strain. Various suggestions concerning
design of novel ferromagnetic semiconductor systems are described.
\end{abstract}


\maketitle

\twocolumn

\section{Introduction}

Because of complementary properties of semiconductor and ferromagnetic
material systems, a growing effort is directed toward studies of
semiconductor-magnetic nanostructures.
Applications in sensors and memories [1] as well as for
computing using electron spins can be envisaged [2]. The hybrid nanostructures, in
which both electric and magnetic field are spatially modulated, are usually fabricated
by patterning of a ferromagnetic metal on the top of a semiconductor [3] or by
incorporation of ferromagnetic clusters into a semiconductor matrice [4]. In such
devices, the stray fields can control charge and spin dynamics in the semiconductor.
At the same time, spin-polarized electrons from the metal could be injected into the
semiconductor.  The efficiency of such a process appears, however, to be prohibitively
low [5-7].

Already the early studies of Cr spinels [8] and rock-salt Eu  [9,10] and Mn-based
[11] chalcogenides led to the observation of a number of outstanding phenomena
associated with the interplay between ferromagnetic cooperative phenomena and
semiconducting properties. The discovery of ferromagnetism in zinc-blende III-V
[12,13] and II-VI [14,15] Mn-based compounds allows one to explore physics of
previously not available combinations of quantum structures and magnetism in
semiconductors. For instance, a possibility of changing the magnetic phase by light in
(In,Mn)As/(Al,Ga)Sb [16] and (Cd,Mn)Te/(Cd,Zn,Mg)Te [14] heterostructures was
put into the evidence. The injection of spin-polarized carriers from (Ga,Mn)As to a
(In,Ga)As quantum well in the absence of an external magnetic field was
demonstrated, too [17]. It is then important to understand the ferromagnetism in these
semiconductors, and to ask whether the Curie temperatures $T_C$ can be raised to above
300 K from the present 110 K observed for Ga$_{0.947}$Mn$_{0.053}$As [13,18].
In this paper, we outline briefly the main ingredients of a model put recently
forward to describe quantitatively the hole-mediated ferromagnetism in tetrahedrally
coordinated semiconductors [19]. We also list a number of counterintuitive
experimental findings, which are explained by the model. Finally, we present the
relevant chemical trends and discuss various suggestions concerning the design of
novel ferromagnetic semiconductor systems. The recent comprehensive reviews
present many other aspects of III-V [20], II-VI [21] as well as of IV-VI [11] magnetic
semiconductors, which are not discussed here.

\section{Origin of ferromagnetism}

Since we aim at quantitative description of experimental findings, the
proposed theoretical approach [19] makes use of empirical facts and parameters
wherever possible. In this section, we discuss those effects, which are regarded as
crucial in determining the magnitude of ferromagnetic couplings in p-type magnetic
semiconductors [19].

\subsection{Charge and spin state of Mn ions}

We consider tetrahedrally coordinated semiconductors, in which the
magnetic ion Mn occupies the cation sublattice, as found by
extended x-ray absorption fine structure (EXAFS) studies in the
case of Cd$_{1-x}$Mn$_x$Te  [22] and Ga$_{1-x}$Mn$_x$As [23]. The
Mn provides a localized spin and, in the case of III-V
semiconductors, acts as an acceptor. These Mn acceptors compensate
the deep antisite donors commonly present in GaAs grown by
low-temperature molecular beam epitaxy, and produce a p-type
conduction with metallic resistance for the Mn concentration $x$
in the range $0.04 \le x \le 0.06$ [18,24-26]. According to optical
studies, Mn in GaAs forms an acceptor center characterized by a
moderate binding energy [27] $E_a = 110$ meV, and a small
magnitude of the energy difference between the triplet and singlet
state of the bound hole [27,28] $\Delta \epsilon = 8 \pm 3$ meV.
This small value demonstrates that the hole introduced by the
divalent Mn in GaAs does not reside on the d shell or forms a
Zhang-Rice-like singlet [29,30], but occupies an effective mass
Bohr orbit [19,31]. Thus, due to a large intra-site correlation
energy $U$, (Ga,Mn)As can be classified as a charge-transfer
insulator, a conclusion consistent with photoemission spectroscopy
[32,33]. At the same time, the p-d hybridization results in a
spin-dependent coupling between the holes and the Mn ions, $H_{pd}
= - \beta N_o\mbox{\boldmath$sS$}$. Here $\beta$ is the p-d
exchange integral and $N_o$ is the concentration of the cation
sites. The analysis of both photoemission data [32,33] and
magnitude of $\Delta \epsilon$ [31] leads to the exchange energy
$\beta N_o \approx -1$~eV. Similar values of $\beta N_o$ are
observed in II-VI diluted magnetic semiconductors with comparable
lattice constants [34]. This confirms Harrison's suggestion that
the hybridization matrix elements depend primarily on the
intra-atomic distance [35]. According to the model in question,
the magnetic electrons remain localized at the magnetic ion, so
that they do not contribute to charge transport. This precludes
Zener's double exchange [36] as the mechanism leading to
ferromagnetic correlation between the distant Mn spins. At the
same time, for some combinations of transition metals and hosts,
the "chemical" attractive potential introduced by the magnetic ion
can be strong enough to bind the hole on the local orbit [29,30].
In an intermediate regime, the probability of finding the hole
around the magnetic ion is enhanced, which result in the apparent
increase of $|\beta N_o|$ with decreasing $x$ [30].

In addition to the carrier-spin interaction, the p-d hybridization leads to the
superexchange, a short-range antiferromagnetic coupling between the Mn spins. In
order to take the influence of this interaction into account, it is convenient to
parameterize the dependence of magnetization on the magnetic field in the absence of
the carriers, $M_o(H)$, by the Brillouin function, in which two empirical parameters, the
effective spin concentration $x_{eff}N_o < xN_o$ and temperature $T_{eff} > T$,
take the presence of
the superexchange interactions into account [34,37,38]. The dependencies $x_{eff}(x)$ and
$T_{AF}(x)\equiv T_{eff}(x) - T$ are known [37,38] for (Zn,Mn)Te,
while $x \approx x_{eff}$ and $T_{AF} \approx 0$ for
(Ga,Mn)As, as explained below.

\subsection{Electronic states near metal-insulator transition}

Because of ionized impurity and magnetic scattering, the effective mass holes
introduced by Mn in III-V compounds or by acceptors such as N or P in the case of
II-VI DMS, are at the localization boundary. It is, therefore, important to discuss the
effect of localization on the onset of ferromagnetism. The two-fluid model [39]
constitutes the established description of electronic states in the vicinity of the
Anderson-Mott metal-insulator transition (MIT) in doped semiconductors. According
to that model, the conversion of itinerant electrons into singly occupied impurity states
with increasing disorder occurs gradually, and begins already on the metal side of the
MIT. This leads to a disorder-driven static phase separation into two types of regions:
one populated by electrons in extended states, and another containing singly occupied
impurity-like states. The latter controls the magnetic response of doped non-magnetic
semiconductors [39] and gives rise to the presence of BMPs on both sides of the MIT
in magnetic semiconductors [34,40,41]. Actually, the formation of BMPs shifts the
MIT towards the higher carrier concentrations [34,40,41]. On crossing the MIT, the
extended states become localized. However, according to the scaling theory of the
MIT, their localization radius $\xi$ decreases rather gradually from infinity at the MIT
towards the Bohr radius deep in the insulator phase, so that on a length scale smaller
than $\xi$ the wave function retains an extended character. Such weakly localized states
are thought to determine the static longitudinal and Hall conductivities of doped
semiconductors.

The central suggestion of the recent model [19] is that the holes
in the extended or weakly localized states mediate the long-range
interactions between the localized spins on both sides of the MIT
in the III-V and II-VI magnetic semiconductors. As will be
discussed below, $T_C$ is proportional to the spin susceptibility
of the carrier gas $\chi_s$ which, in turn, is proportional to the
thermodynamic spin density-of-states $\rho_s$. Like other
thermodynamic quantities, $\rho_s$ does not exhibit any critical
behavior at the MIT. The quantitative renormalization of $\rho_s$
by disorder will depend on its actual form, for instance, on the
degree of compensation. The enhancement of $\rho_s$ by the
carrier-carrier interactions can be described by the Fermi-liquid
parameter $A_F$, $\rho_s \rightarrow A_F\rho_s$.  The value of
$A_F = 1.2$, as evaluated  [42] by the local-spin-density
approximation for the relevant hole concentrations, has been
adopted for the computations.

The participation of the same set of holes in both charge
transport and the ferromagnetic interactions is shown, in
(Ga,Mn)As [24] and in (Zn,Mn)Te [22], by the agreement between the
temperature and field dependencies of the magnetization deduced
from the extraordinary Hall effect, $M_H$, and from direct
magnetization measurements, $M_D$, particularly in the vicinity of
$T_C$. However, below $T_C$ and in the magnetic fields greater
than the coercive force, while $M_H$ saturates (as in standard
ferromagnets), $M_D$ continues to rise with the magnetic field
[24,43]. This increase is assigned to the BMPs, which interact
weakly with the ferromagnetic liquid. To gain the Coulomb energy,
the BMPs are preferentially formed around close pairs of ionized
acceptors. In the case of p-(Ga,Mn)As this leads to a local
ferromagnetic alignment of neighbor Mn d$^5$ negative ions [44], so
that $x \approx x_{eff}$
and $T_{AF} \approx 0$. By contrast, BMPs in
p-(Zn,Mn)Te are not preferentially formed around Mn pairs and encompass more Mn
spins for a given $x$, as the small binding energy [45] $E_a = 54$ meV corresponds to a
relatively large localization radius. The presence of a competition between the
ferromagnetic and antiferromagnetic interactions in II-VI compounds, and its absence
in III-V materials, constitutes the important difference between those two families of
magnetic semiconductors.

\subsection{The structure of the valence band}

The meaningful model of the hole-mediated ferromagnetism has to
take into account the complex nature of the valence band in
semiconductors resulting from $kp$ and spin-orbit interactions.
Therefore, the hole dispersion and wave functions are computed by
diagonalizing the 6x6 Kohn-Luttinger matrix [46] together with the
recalculated p-d exchange contribution [47]. The model is
developed for zinc-blende and wurzite semiconductors, allows for
warping, quantizing magnetic fields, strain, and arbitrary
orientations of $\mbox{\boldmath$M$}$. Because of the spin-orbit
interaction, the exchange splitting of the valence band depends on
the relative orientation of the magnetization and the hole wave
vector. This mixing of orbital and spin degrees of freedom [48]
accounts for substantial differences between the effects of the
exchange interaction upon properties of the electron and hole
liquids. The tabulated values of the effective-mass constants
[49-52] are taken as input parameters. According to interband
magnetoptics [37] and photoemission studies [32] $\beta N_o = -1.1
\pm 0.1$ and $-1.2 \pm 0.2$ eV for (Zn,Mn)Te and (Ga,Mn)As,
respectively.

\subsection{Zener model of carrier-mediated ferromagnetism}

Zener [53] first proposed the model of ferromagnetism driven by the exchange
interaction between carriers and localized spins. However, this model was later
abandoned, as neither the itinerant character of the magnetic electrons nor the
quantum (Friedel) oscillations of the electron spin-polarization around the localized
spins were taken into account, both of these are now established to be critical
ingredients for the theory of magnetic metals. In particular, a resulting competition
between ferromagnetic and antiferromagnetic interactions leads rather to a spin-glass
than to a ferromagnetic ground state. In the case of semiconductors, however, the mean
distance between the carriers is usually much greater than that between the spins.
Under such conditions, the exchange interaction mediated by the carriers is
ferromagnetic for most of the spin pairs, which reduces the tendency towards
spin-glass freezing. Actually, for a random distribution of the localized spins, the
mean-field value of the Curie temperature $T_C$ deduced from the Zener model is equal to that
obtained from the Ruderman, Kittel, Kasuya, and Yosida (RKKY) approach, in which
the presence of the Friedel oscillations is explicitly taken into account [54,55].

The starting point of the model is the determination how the
Ginzburg-Landau free-energy functional $F$ depends on the
magnetization $M$ of the localized spins. As mentioned above, the
hole contribution to $F$, $F_c[M]$ is computed by diagonalizing
the 6x6 Kohn-Luttinger matrix together with the p-d exchange
contribution, and by the subsequent computation of the partition
function $Z$. This model takes the effects of the spin-orbit
interaction into account, a task difficult technically within the
RKKY approach, as the spin-orbit coupling leads to non-scalar
terms in the spin-spin Hamiltonian.  Moreover, the indirect
exchange associated with the virtual spin excitations between the
valence bands, the Bloembergen–-Rowland mechanisms [34], is
automatically included.

The remaining part of the free energy functional, that of the localized spins, is
given by
\begin{equation}
F_S[M] = \int_0^M d M_o H(M_o),         (1)
\end{equation}
where $H(M_o)$ is the inverse function of the experimental dependence of the
magnetization on the magnetic field H in the absence of the carriers. By minimizing $F =
F_c[M] + F_S[M]$ with respect to $M$ at given $T$, $H$, and hole concentration $p$,
one obtains
$M(T,H)$ within the mean-field approximation, an approach that is quantitatively valid
for long-range exchange interactions.

\subsection{Curie temperature}

As described above, $T_C$ can be computed by minimizing the free
energy, and without referring to the explicit form of the carrier
periodic wave functions $u_{i\mbox{\boldmath$k$}}$. Since near
$T_C$ the relevant magnetization $M$ becomes small, the carrier
free energy, and thus $T_C$, can also be determined from the
linear response theory. The resulting $T_C$ assumes the general
form
\begin{equation}
T_C = x_{eff}N_oS(S+1)\beta^2A_F\rho_s(0,T_C)/12k_B - T_{AF}.
\end{equation}
In the absence of disorder
\begin{equation}
\rho_s(\mbox{\boldmath$q$},T) = 8\sum_{ij\mbox{\boldmath$k$}}
\frac{|\langle
u_{i\mbox{\boldmath$k$}}|s_z|u_{j\mbox{\boldmath$k$}+\mbox{\boldmath
$q$}}\rangle|^2 f_i(\mbox{\boldmath$k$})[1 - f_j(\mbox{\boldmath
$k$}+\mbox{\boldmath$q$})]}
{[E_j(\mbox{\boldmath$k$}+\mbox{\boldmath$q$})-
E_i(\mbox{\boldmath$k$})]},
\end{equation}
where $z$ is the direction of magnetization and
$f_i(\mbox{\boldmath$k$})$ is the Fermi-Dirac distribution
function for the $i$-th valence band subband. It is seen that for
spatially uniform magnetization (the ferromagnetic case), only the
terms corresponding to $q = 0$ (i.e., the diagonal terms)
determine $T_C$. For the strongly degenerate carrier gas as well
as neglecting the spin-orbit interaction, $\rho_s$ is equal to the
total density-of-states $\rho$ for intra-band charge excitations,
where $\rho = m^*_{DOS}k_F/\pi^2\hbar^2$.

Equation (2) for $T_C$ with $\rho_s = \rho$ has already been
derived by a number of equivalent methods [34,42,54,55]. It is
straightforward to generalize the model for the case of
the carriers confined to the $d$-dimensional space [14,54,55].
The tendency towards the formation of
spin-density waves in low-dimensional systems [55] as well as possible
spatial correlation in the distribution of the
magnetic ions can also be taken into account.
The mean-field value of the critical temperature
$T_{\mbox{\boldmath$q$}}$, at which the system undergoes the
transition to a spatially modulated state characterized by the
wave vector $\mbox{\boldmath$q$}$, is given by the solution of the
equation,
\begin{equation}
\beta^2A_F(\mbox{\boldmath$q$},T_{\mbox{\boldmath
$q$}})\rho_s(\mbox{\boldmath$q$},T_{\mbox{\boldmath$q$}}) \int d
\mbox{\boldmath$\zeta$}\chi_o(\mbox{\boldmath$q$},T_{\mbox{\boldmath
$q$}},\mbox{\boldmath$\zeta$}) |\phi_o(\mbox{\boldmath$\zeta$})|^4
= 4g^2\mu_B^2.
\end{equation}
Here $\mbox{\boldmath$q$}$ spans the $d$-dimensional space,
$\phi_o(\mbox{\boldmath$\zeta$})$ is the envelope function of the
carriers confined by a $(3 - d)$-dimensional potential well
$V(\mbox{\boldmath$\zeta$})$; $g$ and $\chi_o$ denote the Land\'e
factor and the $\mbox{\boldmath$q$}$-dependent magnetic
susceptibility of the magnetic ions in the absence of the
carriers, respectively. Within the MFA, such magnetization shape
and direction will occur in the ordered phase, for which the
corresponding $T_{\mbox{\boldmath$q$}}$ attains the highest value.

\section{Comparison of the model to selected experimental results}
\subsection{Magnetic circular dichroism in (Ga,Mn)As}

In the case of II-VI DMS, detail information on the exchange-induced
spin-splitting of the bands, and thus on the coupling between the effective mass electrons
and the localized spins has been obtained from magnetooptical studies [34,47]. A
similar work on (Ga,Mn)As [43,56,57] led to a number of surprises. The most striking
was the opposite order of the absorption edges corresponding to the two circular
photon polarizations in (Ga,Mn)As comparing to II-VI materials. This behavior of
circular magnetic dichroism (MCD) suggested the opposite order of the
exchange-split spin subbands, and thus a different origin of the sp-d interaction in these two
families of DMS. A new light on the issue was shed by studies of photoluminescence
(PL) and its excitation spectra (PLE) in p-type (Cd,Mn)Te quantum wells [14].  As
shown schematically in Fig. 1, the reversal of the order of PLE edges corresponding to
the two circular polarizations results from the Moss-Burstein effect, that is from the
shifts of the absorption edges associated with the empty portion of the valence
subbands in the p-type material. This model was subsequently applied to interpret
qualitatively the magnetoabsorption data for metallic (Ga,Mn)As [57]. Surprisingly,
however, the anomalous sign of the MCD was present also in non-metallic
(Ga,Mn)As, in which EPR signal from occupied Mn acceptors was seen [57]. It has,
therefore, been suggested that the exchange interaction between photo- and
bound-holes is responsible for the anomalous sign of the MCD in those cases [57]. The
presence of such a strong exchange mechanism is rather puzzling, and it should
be seen in non-magnetic p-type semiconductors. At the same time,
according to our two-fluid model, the co-existence
of strongly and weakly localized holes is actually expected on the both sides of the
MIT. Since the Moss-Burstein effect operates for interband optical transitions
involving weakly localized states, it leads to the sign reversal of the MCD, also on the
insulating side of the MIT.

\begin{figure}
\includegraphics*[width=80mm]{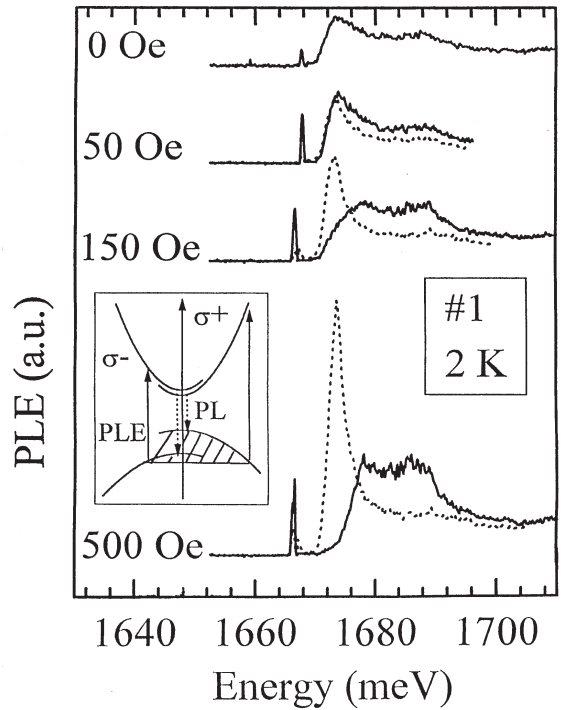}
 \caption{Photoluminescence excitation spectra (PLE), that is the
photoluminescence (PL) intensity as a function of the excitation
photon energy intensity, for $\sigma^+$ (solid lines) and
$\sigma^-$ (dotted lines) circular polarizations at selected
values of the magnetic field in a modulation-doped p-type quantum
well of Cd$_{0.976}$Mn$_{0.024}$Te at 2 K. The photoluminescence
was collected in $\sigma^+$ polarization at energies marked by the
narrowest features. The sharp maximum and step-like form
correspond to quasi-free exciton and transitions starting at the
Fermi level, respectively. Note reverse ordering of transition
energies at $\sigma^+$ and $\sigma^-$  for PL and PLE (the latter
is equivalent to optical absorption). The band arrangement at 150
Oe is sketched in the inset (after [14]).}
\end{figure}

Another striking property of the MCD is a different temperature dependence of
the normalized MCD at low and high photon energies in ferromagnetic (Ga,Mn)As
[43]. This observation was taken as an evidence for the presence of two spectrally
distinct contributions to optical absorption [43]. A quantitative computation of MCD
spectra has recently been undertaken [58]. The theoretical results demonstrate that
because of the Moss-Burstein effect, the magnetization-induced splitting of the bands
leads to a large energy difference between the positions of the absorption edges
corresponding to the two opposite circular polarizations. This causes an unusual
dependence of the low-energy onset of MCD on magnetization, and thus on
temperature. These considerations lead to a quantitative agreement with the
experimental findings, provided that the actual hole dispersion and wave functions are
taken for the computation of MCD.

\subsection{Curie temperature in (Ga,Mn)As, (Zn,Mn)Te and (Cd,Mn)Te quantum
well}

The most interesting property of Ga$_{1-x}$Mn$_x$As epilayers is the large magnitude
of $T_C$, of the order of 100 K for the Mn concentration $x$ as low as 5\% [13,18]. Because
of this high $T_C$, the spin-dependent extraordinary contribution to the Hall resistance $R_H$
persists up to 300 K, making an accurate determination of the hole density difficult
[18, 24-26]. However, the recent measurement [59] of $R_H$ up to 27 T and at 50 mK
yielded an unambiguous value of $p = 3.5\times10^{20}$ cm$^{-3}$
for a metallic Ga$_{0.947}$Mn$_{0.053}$As
sample, in which $T_C$ = 110 K is observed  [18]. As shown in Fig. 2, the present model
explains, with no adjustable parameters, such a high value of $T_C$.

\begin{figure}
\includegraphics*[width=80mm]{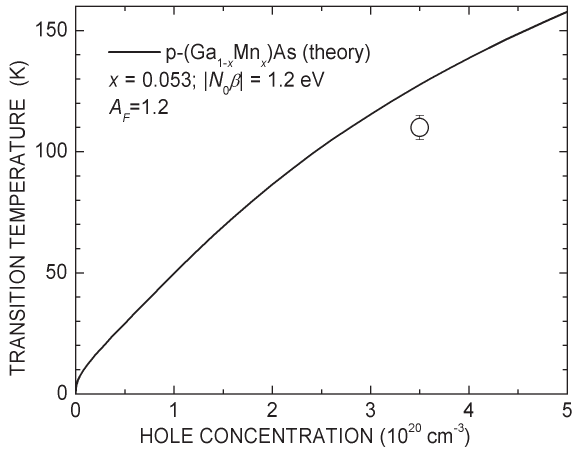}
\caption{Calculated ferromagnetic transition temperature as a
function of hole concentration for Ga$_{1-x}$Mn$_x$As with $x$ =
0.053 [19]. The open circle indicates the experimental result for $p
= 3.5\times 10^{20}$ cm$^{-3}$ (after [18]).}
\end{figure}

The studied epilayers of (Zn,Mn)Te:N [15] were on the insulating
side of the metal-insulator transition (MIT), as the bound
magnetic polaron (BMP) formation enhances localization.
Nevertheless, if the concentration of acceptors was sufficiently
high, the ferromagnetic Curie-Weiss temperatures $T_{CW}$ were
observed as well as magnetic hysteresis were detected below $T_C
\approx T_{CW}$ [15]. At the same time, the values of $T_C$ were
much lower than those characterizing (Ga,Mn)As. However, a
comparison of the experimental and calculated values of $T_C$ for
(Zn,Mn)Te as a function of $x$ and $p$ (Fig. 3) demonstrates that
the present model is capable to explain the magnitude of $T_C$
except for the samples with the smallest $x$. In the latter,
$p/xN_o$ is as large as 0.6, so that precursor effects of Friedel
oscillations and Kondo correlation are expected at low
temperatures  [54].

\begin{figure}
\includegraphics*[width=80mm]{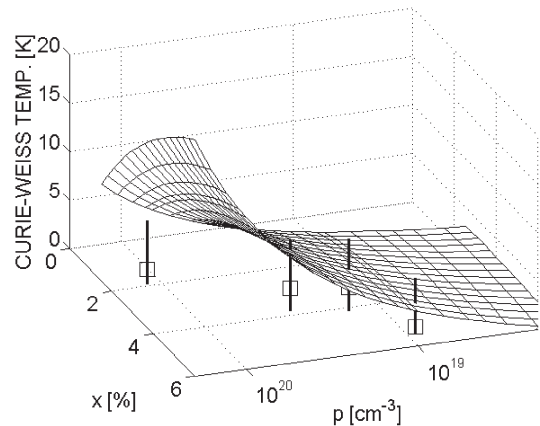}
\caption{Curie-Weiss temperature in Zn$_{1-x}$Mn$_x$Te:N for
various Mn contents $x$ and hole concentrations $p$, determined
from the temperature dependence of the magnetic susceptibility.
Experimental values are marked by squares [15] while theoretical
predictions by the mesh (after [19]).}
\end{figure}

Two effects appear to account for the greater $T_C$ values in
p-(Ga,Mn)As than in p-(Zn,Mn)Te at given $p$ and $x$. First is the
smaller magnitude of the spin-orbit splitting between the
$\Gamma_8$ and $\Gamma_7$ bands in arsenides, $\Delta_o = 0.34$
eV, in comparison to that of tellurides, $\Delta_o = 0.91$ eV.
Once the Fermi energy $E_F$ approaches the $\Gamma_7$ band, the
density-of-states effective mass increases, and the reduction of
the carrier spin susceptibility by the spin-orbit interaction is
diminished. The computed value of $T_C$ for $p = 3\times 10^{20}$
cm$^{-3}$ is greater by a factor of four in (Ga,Mn)As than that
evaluated in the limit $\Delta_o >> E_F$. The other difference
between the two materials is the destructive effect of
antiferromagnetic interactions, which operate in II-VI compounds
but are of minor importance in III-V materials, as explained in
Sec.~2.

The model discussed above describes also the magnitude of $T_C$ and its
dependence on $x$ in modulation-doped quantum wells of p-(Cd,Mn)Te, if $A_F = 2$ is
assumed [60], an expected value for the relevant densities of the two-dimensional hole
liquid. A good description of $T_C(p)$ is also obtained, provided that disorder broadening
of the density-of-states at low $p$ is taken into account [60]. Whether the ground state
corresponds to uniform magnetization or rather to a spin-density wave is under study
now.

\subsection{Effects of strain}

Already early studies of a ferromagnetic phase in (Ga,Mn)As epilayers
demonstrated the existence of substantial magnetic anisotropy [61]. Magnetic
anisotropy is usually associated with the interaction between spin and orbital degrees
of freedom of the magnetic electrons. According to the model in question, these
electrons are in the d$^5$ configuration. For such a case the orbital momentum $L = 0$, so
that no effects stemming from the spin-orbit coupling could be expected.  To reconcile
the model and the experimental observations, we note that the interaction between the
localized spins is mediated by the holes, characterized by a non-zero orbital
momentum. An important aspect of the present model is that it does take into account
the anisotropy of the carrier-mediated exchange interaction associated with the
spin-orbit coupling in the host material, an effect difficult to include within the standard
approach to the RKKY interaction.

The computed effect of the cubic anisotropy on $T_C$ has been found to be small;
differences between $T_C$ values calculated for various orientations of magnetization in
respect to crystallographic axes are below 0.1 K in (Ga,Mn)As [58]. The
corresponding differences are, however, greater in the presence of epitaxial strain; of
the order of 1 K for 1\% biaxial strain in the (001) plane. Thus, such a strain can
control the orientation of the easy axis. According to the computation for the relevant
hole concentrations, the easy axis is in the plane for the case of unstrained or
compressively strained films but under tensile strain the easy axis takes the [001]
direction. These expectations [58] are corroborated by the experimental study, in
which appropriate substrates allowed to control the direction of strain [61]. It worth
noting that similarly to strain, also confinement of the holes affects the magnetic
anisotropy – in accord with the theoretical model, the easy axis is oriented along the
growth direction in the ferromagnetic p-(Cd,Mn)Te quantum wells [14,60].

\section{Other perspective materials}

In view of the general agreement between experiment and theory for $T_C$ and
the magnetic anisotropy, it is tempting to extend the model for material systems that
might be suitable for fabrication of novel ferromagnetic semiconductors. For instance,
the model suggests immediately that $T_C$ values above 300 K could be achieved in
Ga$_{0.9}$Mn$_{0.1}$As, if such a large value of $x$ would be accompanied by a corresponding
increase of the hole concentration.
Figure 4 presents the values of $T_C$ computed for various tetrahedrally coordinated
semiconductors containing 5\% of Mn and $3.5\times 10^{20}$ holes per cm$^3$ [19].
In addition to
adopting the tabulated values of $\gamma_i$ and $\Delta_o$ [49-52] the same value of $\beta =
\beta$[(Ga,Mn)As] for all group IV and III-V compounds was assumed, which results in
an increase of $|\beta N_o| \sim a_o^{-3}$,
where $a_o$ is the lattice constant, a trend known to be obeyed
within the II-VI family of magnetic semiconductors [34]. By extending the model for
wurzite semiconductors, $T_C$ values for parameters of ZnO [62] (Fig. 3) and for wurzite
GaN (not shown) were evaluated. For the employed parameters [63] the magnitude of
$T_C$ for the cubic GaN (Fig. 4) is by 6\% greater than that computed for the wurzite
structure.

\begin{figure}
\includegraphics*[width=80mm]{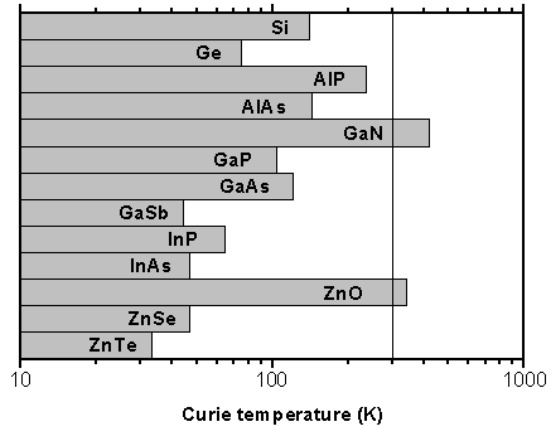}
\caption{Computed values of the Curie temperature $T_C$ for
various p-type semiconductors containing 5\% of Mn and $3.5\times
10^{20}$ holes per cm$^3$ (after [19]).}
\end{figure}

The data (Fig. 4) demonstrate that there is much room for a
further increase of $T_C$ in p-type magnetic semiconductors. In
particular, a general tendency for greater $T_C$ values in the
case of lighter elements stems from the corresponding increase in
p-d hybridization and reduction of spin-orbit coupling. It can be
expected that this tendency is not altered by the uncertainties in
the values of the relevant parameters. Important issues of
solubility limits and self-compensation as well as of the
transition to a strong-coupling case with decreasing $a_o$ [30]
need to be addressed experimentally. We note in this context that
since, in general, III-V compounds can easier be doped by
impurities that are electrically active, whereas II-VI materials
by transition metals, a suggestion has been put forward to grow
magnetic III-V/II-VI short period superlattice [64].

Finally, we address the important question about the feasibility
of synthesizing n-type or intrinsic ferromagnetic semiconductors.
A work on (Zn,Mn)O:Al [65] is relevant in this context. One should
not forget, however, about the existence of, e.g., europium
chalcogenides and chromium spinels, whose ferromagnetism is not
driven by free carriers. Actually, a theoretical suggestion has
been made [66] that superexchange in Cr-based II-VI compounds can
lead to a ferromagnetic order. Desired material properties, such
as divergent magnetic susceptibility and spontaneous
magnetization, can also be achieved in the case of a strong
antiferromagnetic super-exchange interaction.  The idea here [67]
is to synthesize a ferrimagnetic system that would consist of
antiferromagnetically coupled alternating layers containing
different magnetic cations, e.g., Mn and Co.

The above list of possibilities is by no means exhausting. With no doubt we
will witness many unforeseen developments in the field of ferromagnetic
semiconductors in the near future.

\section*{Acknowledgments}

The work at Tohoku University was supported by the Japan Society
for the Promotion of Science and by the Ministry of Education,
Japan; the work in Poland by State Committee for Scientific
Research, Grant No. 2-P03B-02417, and by Foundation for Polish
Science. We are grateful to F. Matsukura and Y. Ohno for
collaboration on III-V magnetic semiconductors. One of us (T~D.)
thanks J.~Jaroszy\'nski and P.~Kossacki in Warsaw and J.~Cibert
and D.~Ferrand in Grenoble for collaboration on II-VI magnetic
semiconductors.



\end{document}